\documentclass[doublespacing]{elsart}
\usepackage[english]{babel}

\usepackage{epsfig}
\usepackage{graphics}

\usepackage{amssymb}

\begin{document}

\begin{frontmatter}

\title{Relativistic Split-Cavity Oscillator}

\author{V.G. Baryshevsky}

\address{Research Institute for Nuclear Problems, Belarusian State
University, 11~Bobruiskaya Str., Minsk 220030, Belarus}
\ead{bar@inp.bsu.by, v\_baryshevsky@yahoo.com}

\begin{abstract}
Using the method of small signal analysis, we study the
application potential of relativistic electron beams in
split-cavity oscillators (SCO's). A beam-energy change in the SCO
as a function of the initial energy of a relativistic beam is
considered. It is shown that the small-signal analysis method
enables adequate evaluation of SCO parameters needed for effective
modulation of a relativistic beam  in a split cavity and for HPM
generation using SCO's. The range of energies is found for which
the  effect of self-modulation of the beam density in SCO
structures is most pronounced. It is also shown that for beam
currents at which the space charge has little effect on the
 motion of electrons in a beam, the beam in a split-cavity oscillator is effectively
self-modulated at beam energies less than $\approx 300 \div
400$\,keV. The self-modulation drops sharply in the range of
energies from 250 to 400 keV, but as the beam current is
increased, the effective beam self-modulation becomes appreciable
in this range too, as well as even in a higher energy range.
\end{abstract}

\end{frontmatter}

\section{Introduction}
 The study of radiation generation mechanisms via the
transit-time effect is one of the lines of research into the
 process of high-power microwave generation. Transit-time
effect oscillators (TTO) (monotron \cite{marc,barr00,barr01},
split-cavity oscillator (SCO) \cite{mard}, super-reltron
\cite{mil}, double-foil SCO \cite{lemke,he}) enabled  the HPM
generation using high-current electron beams without the need for
an applied magnetic field.

In a fundamental work \cite{mard}, Marder and colleagues not only
gave a detailed computer simulation of SCO behavior, but also
developed in the case of small signal analysis a method allowing
the evaluation of  SCO parameters required for effective
modulation of nonrelativistic electron beam  in SCO's (for
effective HPM generation through the use of SCO extractors).

In view of the extensive research into the use of high-current
relativistic electron beams for HPM generation, in this paper we
shall generalize the small signal analysis method  suggested in
\cite{mard} to the case of  relativistic beams passing through a
split cavity.

It is shown that the small signal analysis method enables adequate
evaluation of  SCO parameters needed for effective modulation of a
relativistic beam  in a split cavity and for HPM generation using
SCO's. The range of energies is found for which the  effect of
self-modulation of the beam density in SCO structures is most
pronounced.

\section{Small Signal Analysis of the SCO}

The split-cavity oscillator (SCO) \cite{mard} consists of a high-Q
pillbox cavity with conducting screen walls through which an
electron beam can pass. The cavity is symmetrically partitioned by
a screen which leaves a gap between it and the outer cavity wall
(see Figure 1).

\begin{figure}[htbp]
\label{fig1}
\begin{center}
     \resizebox{35mm}{!}
       {\includegraphics{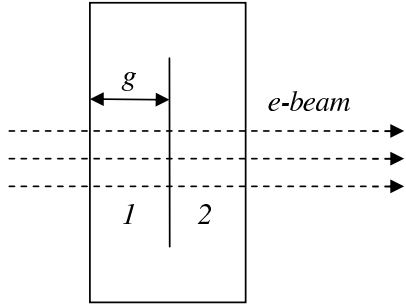}}\\
\caption{SCO geometry}
\end{center}
\end{figure}

 Following Marder and co-authors \cite{mard}, to describe the behavior of the SCO in the case of relativistic
 electron beams, we shall first consider the transit-time oscillator (see Figure 2) excited by a relativistic beam.

 The TTO pillbox is like an SCO,  but without partition (see Figure 2). According to  \cite{mard},
 the pillbox cavity of the TTO can be approximated by a one-dimensional gap of width $g$ across which a uniform RF electric field of the form
 \begin{equation}
 \label{eq1}
 \vec E=\vec{E}_0\sin (\omega\, t +\vartheta)
 \end{equation}
 is imposed.

\begin{figure}[htbp]
\label{fig1}
\begin{center}
       \resizebox{35mm}{!}
       {\includegraphics{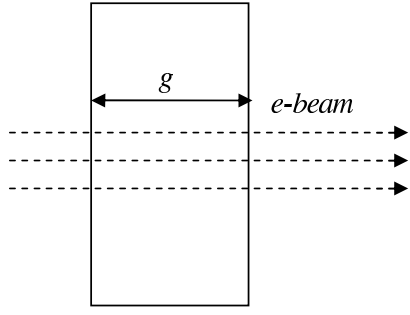}}\\
\caption{TTO geometry}
\end{center}
\end{figure}

 It is assumed that the electric field $\vec E$ is directed
along the $z$-axis and the electron entering the pillbox cavity
moves parallel to the $z$-axis; $\omega$ is the field frequency
and $\vartheta$ is the initial phase of the field (the phase of
the field at the moment when the particle enters the pillbox
cavity). The magnetic field $\vec H$ exists in the pillbox cavity
together with the electric field. However, in the case of small
signals analysis, when the Larmor radius of particle motion in a
magnetic field is much greater than the width $g$, the magnetic
field only causes the displacement of the electron in the
direction orthogonal to the $z$-axis, but has no effect on the
longitudinal motion of the particle along the $z$-axis.  In what
follows we shall neglect the influence of the magnetic field on
particle motion. As a result, the equation of motion along the
$z$-axis for the electron entering into the pillbox cavity can be
written in the form:
\begin{equation}
\label{eq2} \frac{dp}{dt}=-e\,E(t)=-e\,E_0\sin(\omega t+\vartheta)
\end{equation}
where $ p$ is the momentum of the electron and $e$ is the value of
the electron charge; $p = m\gamma v$, where $m$ is the mass of the
electron and $\gamma=\frac{1}{\sqrt{1-\frac {v^2}{c^2}}}$ is the
Lorentz factor, with $v$ and c being the electron velocity and the
speed of light, respectively.

It follows from (\ref{eq2}) that
\begin{equation}\label{eq3}
p(t,v)=p_0+\frac{e E_0}{\omega}[\cos(\omega
t+\vartheta)-\cos\vartheta],
\end{equation}
that is,
\begin{eqnarray}
\label{eq4} \gamma
v & = & \frac{v}{\sqrt{1-\frac{v^2}{c^2}}}\\
& = &
\gamma_0v_0\left\{1+\frac{eE_0}{m\gamma_0v_0\omega}[\cos(\omega
t+\vartheta)-\cos\vartheta]\right\}=F(t,\vartheta),\nonumber
\end{eqnarray}
where $\gamma_0$, $v_0$ are the particle's Lorentz factor and
velocity at the moment of entering the pillbox cavity,
respectively.

The solution of (\ref{eq4}) has the form
\begin{equation}
\label{eq5}
v=\frac{dz}{dt}=\pm\frac{F}{\sqrt{1+\frac{1}{c^2}F^2}}.
\end{equation}
We shall further be interested  in the motion in the positive
direction of the $z$-axis, so in (\ref{eq5}) we shall only
consider the solutions with the plus sign. Equation (\ref{eq5})
yields the following expression for the trajectory of
relativistic-particle motion in a pillbox cavity
\begin{equation}
\label{eq6} z(t)=\int^t_0F(t^{\prime},
\vartheta)\frac{1}{\sqrt{1+\frac{1}{c^2}F(t^{\prime},
\vartheta)}}dt^{\prime}
\end{equation}
Let $T$ denote the time when the electron crosses the right border
of the pillbox cavity (Figure 2) (T - is the moment when the
electron escapes from the cavity). According to (\ref{eq6}), we
have
\begin{equation}
\label{eq7} \int^T_0 F(t^{\prime},
\vartheta)\frac{1}{\sqrt{1+\frac{1}{c^2}F(t^{\prime},
\vartheta)}}dt^{\prime}=g
\end{equation}
Using Eqs. (\ref{eq6}), (\ref{eq7}), we can find the time  needed
for the particle  to pass the distance $g$ in the presence of  the
field $E$.

The function $F$ depends on the parameter
\[
\varepsilon=\frac{e E_0}{m\gamma_0 v_0\omega}.
\]
In the case of small signal analysis, the parameter
$\varepsilon\ll 1$. Let us expand the integrand in (\ref{eq7})
into a power series in the small parameter $\varepsilon$. The
left-hand side of (\ref{eq7}) can be integrated within an accuracy
up to the terms proportional to the parameter $\varepsilon$ raised
to the first power, which gives the following expression for the
time $T$:
\begin{equation}
\label{eq8} T=\frac{g}{{v}_0}-{\tilde{\varepsilon}}
\frac{[\sin(\omega {T}_0+\vartheta)-\sin\vartheta-\omega
{T}_0\cos\vartheta]}{\omega},
\end{equation}
where we introduced the following notation:
\[
\tilde{\varepsilon}=\frac{1}{\gamma^2}{\varepsilon},\quad
\varepsilon=\frac{e\, E_0}{m\gamma_0 v_0\omega},\quad
{T}_0=\frac{g}{{v}_0}.
\]
In a nonrelativistic case, when $\frac{v_0}{c}\ll 1$, expression
(\ref{eq8}) goes over to  expression (4) in \cite{mard}.

With (\ref{eq8}) we can find the momentum and then the energy $W$
 of the particle at the moment when it crosses the right border of
the pillbox cavity (Figure 2),  as well as the difference $W-W_0$
between the final  energy $W$ of the particle when it leaves the
pillbox cavity and its initial energy $W_0$ when it enters the
cavity.

We shall consider the difference $W^2-W_0^2$ because the relation
{${W^2=p^2 c^2 + m^2 c^4}$} between $W^2$ and the squared momentum
$p^2$ of the particle is more convenient to use in our further
analysis.

For this difference averaged over the initial phase $\vartheta$ of
particle entrance into the cavity, we can obtain the following
expression:
\begin{equation}
\label{eq9} \frac{W^2-W_0^2}{p_0^2\,
c^2}=\frac{1}{2\pi}\int_0^{2\pi}\left(\frac{p^2(T,\vartheta)}{p_0^2}-1\right)d\vartheta.
\end{equation}
After some cumbersome calculations, we can recast (\ref{eq9}) in
the form:
\begin{equation}
\label{eq10}
\frac{E^2-E_0^2}{p_0^2\,c^2}=(\varepsilon\tilde{\varepsilon}+\varepsilon^2)(1-\cos\omega\,{T}_0)-\varepsilon\tilde{\varepsilon}
\omega \,T_0\sin\omega \,T_0,
\end{equation}
i.e.,
\begin{equation}
\label{eq10a} \frac{E^2-
E^2_{0}}{p_0^2\,c^2\,\varepsilon^2}=\frac{\gamma^2+1}{\gamma^2}(1-\cos\omega\,
T_0)-\frac{1}{\gamma^2}\omega \,T_0\sin \omega \, T_0.
\end{equation}
 In a nonrelativistic case, $\gamma^2\rightarrow 1$, and we have the same result as in \cite{mard}
\begin{equation}
\label{eq11}
\frac{E^2-E_0^2}{p_0^2\,c^2\,\varepsilon^2}=\frac{K-K_0}{K\varepsilon^2}=
2(1-\cos\omega\,{T}_0)-\omega\,{T}_0\sin \omega\,{T}_0,
\end{equation}
 where $K$ is the kinetic energy of the particle at time $T$,
$K_0$ is the kinetic energy of the particle at entering the
cavity.

Now let us consider the split-cavity oscillator (Figure 1). In the
case of small signal analysis,  the alternating  electric field in
the SCO's cavity  can be written as \cite{mard}
\begin{eqnarray}
\label{eq12} E=E_0\sin(\omega t+\vartheta)\quad\mbox{for}\quad
0\leq z < g,\nonumber\\
E=-E_0\sin(\omega t+\vartheta)\quad\mbox{for}\quad g\leq x\leq 2g.
\end{eqnarray}
Using (\ref{eq12}), from the equation of motion similar to
(\ref{eq2}), we can obtain the following expression for particle's
momentum and velocity in an SCO for the time interval $t$ between
the moments when the particle enters the SCO  and when it crosses
the foil placed at point $z=g$, i.e., when the inequality
{{${0\leq t\leq T_g}$ }} holds (here $T_g$ is the moment of time
when the particle crosses the foil):
\begin{eqnarray}
\label{eq13} \gamma \,v = F_{1SCO}(t, \vartheta)=\gamma_0\,
v_0\{1+\varepsilon([\cos(\omega t+\vartheta)-\cos\vartheta])\}
\end{eqnarray}

After the  particle enters the second region of the SCO, the
expression for particle's momentum and velocity for the time
interval $T_g\leq t\leq T$ ($T$ is the moment of time when the
particle leaves the SCO structure, or the time needed for the
particle to pass through the SCO) has the form:
\begin{eqnarray}
\label{eq14} \gamma v  & =&   F_{2SCO}(t,\vartheta)
\\ & = & \gamma_0
v_0 \{1  +  \varepsilon([\cos(\omega\,
T_g+\vartheta)-\cos\vartheta] - [\cos(\omega t+\vartheta)-\cos
(\omega \,T_g+\vartheta)]\},\nonumber
\end{eqnarray}

Equations (\ref{eq13}) and (\ref{eq14}) yield the equations
defining the particle's trajectory $z(t)$ that coincide in form
with (\ref{eq6}) with the function $F(t,\vartheta)$ replaced by
$F_{SCO}(t,\vartheta)$. As a result, the equation defining the
time  $T$ needed for the particle to traverse the area occupied by
the SCO has the form:
\begin{equation}
\label{eq15} \int^T_0
F_{SCO}(t^{\prime},\vartheta)\frac{1}{\sqrt{1+\frac{1}{c^2}F^2(t^{\prime},
\vartheta)}}dt^{\prime}=2g.
\end{equation}
The function $F_{SCO}$ can be written in the form:
\begin{eqnarray*}
& &F_{SCO}=F_{1SCO}\qquad \mbox{for the time $t$ in the
interval}\quad 0\leq t < T_g,\\
& & F_{SCO}=F_{2SCO}\qquad \mbox{for the time $t$ in the interval}
\qquad T_g\leq t\leq T.
\end{eqnarray*}

 In the linear
in $\varepsilon$ approximation, (\ref{eq15}) yields the follows
expression for time~ $T$
\begin{eqnarray}
\label{eq16} T & = &
\frac{2g}{{v}_0}-\tilde{\varepsilon}\frac{[\sin(\omega
{T}_{g_0}+\vartheta)-\sin
\vartheta-\omega{T}_{g_0}\cos\vartheta]}{\omega}
\\
& + & \frac{\tilde{\varepsilon}}{\omega} [\sin(\omega
{T}_0+\vartheta)-\sin\vartheta(\omega{T}_{0g}+ \vartheta) -
\omega{T}_{g0}[2\cos(\omega\,{T}_{g0}+\vartheta)-
\cos\vartheta]],\nonumber
\end{eqnarray}
where
\[
{T}_0=\frac{2g}{{v}_0},\qquad {T}_{g0}=\frac{g}{{v}_0}.
\]
 As a result, using
(\ref{eq14}),  we can obtain the expression $p(T) = m\gamma (T) v
(T)$ for particle momentum at time $T$ and then, the following
expression for the energy change due to the interaction of a
relativistic particle and the field $E(t)$ in a SCO structure
\begin{eqnarray}
\label{eq17} & & \frac{E^2-E_0^2}{p^2_0 c^2}  =
\frac{1}{2\pi}\int\limits^{2\pi}_0 \left(\frac{\gamma^2
v^2}{\gamma_0^2 v^2_0}-1\right) d\vartheta =
\varepsilon\tilde{\varepsilon}(3-4\cos\omega T_{g0}- 4
\omega T_{g0}\,\sin \omega T_{g0}\nonumber\\
& & +\cos 2\omega T_{g0}+ 2 \omega T_{g0}\sin 2\omega T_{g0})  +
\varepsilon^2(3-4\cos\omega T_{g0}+\cos 2\omega T_{g 0}),
\end{eqnarray}
i.e.,
\begin{eqnarray}
\label{eq18} \frac{E^2-E_0^2}{p_0^2\,c^2\,\varepsilon^2} & = &
\frac{1+\gamma^2}{\gamma^2}(3-4\cos
\omega\,T_{g0}+\cos2\omega\,T_{g0}) \\
& - & \frac{1}{\gamma^2}(4\omega \, T_{g0}\sin \omega\, T_{g0}-
2\omega \, T_{g0}\sin 2\omega\, T_{g0})\nonumber.
\end{eqnarray}
 In a nonrelativistic case ($\gamma^2\rightarrow 1$), for the
energy change we have the  expression derived in  \cite{mard}
\begin{eqnarray}
\label{eq19} \frac{E^2-E_0^2}{p_0^2 c^2 \varepsilon^2} & = &
\frac{K-K_0}{K_0
\varepsilon^2}\\
& = & \left[6-8 \cos \omega T_{g0}+2 \cos 2 \omega T_{g0} -  4
\omega T_{g0} \sin \omega T_{g0} + 2 \omega T_{g0} \sin 2 \omega
T_{g0}\right].\nonumber
\end{eqnarray}

Expressions (\ref{eq10a}) and (\ref{eq18}) enable the analysis of
the energy change in  TTO  and  SCO  structures, depending on the
parameters of the system for a relativistic case.

The curves in Fig. 3 show the characteristic beam instability
region {$\left(\frac{E^2-E_0^2}{p_0^2c^2\varepsilon^2} <
0\right)$} when the transit time $T_{g0}$ is about a quarter
period \cite{mard}. However,  in contrast to a nonrelativistic
case, the depth of the instability region decreases as the
particle energy is increased ($\gamma$ is increased), and the
instability drops sharply when the energy of electrons reaches the
range from 250 to 400 \,keV.
 The instability region remains only for large transit
times (right minimum).
\begin{figure}[htbp]
\label{fig3}
\begin{center}
       \resizebox{75mm}{!}{\includegraphics{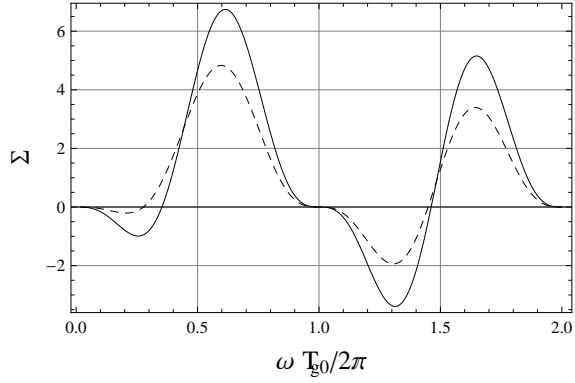}}\\
\caption{Energy change $\left(\Sigma=\frac{E^2- E_0^2}{p_0^2
c^2\varepsilon^2\,\omega\,T_{g0}}\right)$ versus the transit time
$T_{g0}$ for the initial  energies $E_0= 50$\,keV (solid) and
$E_0=200$\,keV (dashed).}
\end{center}
\end{figure}

 But note must be taken of the fact that
this result is valid until the beam density is such that the
effect of space charge on this type of instability can be
neglected.
According to \cite{mard}, as  the current of a nonorelativistic
beam is increased, the space charge effect on instability
enhances. Our analysis shows that in the case of relativistic
beams, the effect of space charge on this type of instability also
enhances with growing beam current.  As a result, the beam
self-modulation effect in SCO's occurs at energies greater than
250\,keV too.
 This can be explained by the fact that when
 the beam density in SCO (TTO) structures rises, the beam instability grows,
 because the current in the beam approaches the range of values at which the vircator can be formed and oscillations
 in the vircator can be excited. As a result, the relativistic beam escaping the SCO appears
 to be modulated within a wide range of currents and space charge densities and their values can
 even be below the threshold for vircator formation.

 It is well known that  a charged particle
 passing through the medium-vacuum boundary emits transition radiation
 \cite{landau}. According to \cite{baryshevsky}, the number of transition radiation
 quanta, $N_{ph}$, produced by a modulated beam can be written in
 the form:
 \begin{equation}
 \label{eq20}
 N_{ph}=N_e\left(N_{1ph}+\left. c\frac{d\,^2 N_{1ph}}{d\omega
 d\Omega}\right|_{\vec K = \vec{\tau}}\,\frac{\pi}{2}\mu^2\frac{\rho_0}{k_0^2}\right),
 \end{equation}
where $N_e$ is the number of electrons that have passed through
the medium-vacuum boundary, $N_{1ph}$ is the number of quanta
produced by one electron, $k_0$ is the photon wave number, and
$\mu=\frac{\rho_1}{\rho_0}$. It is assumed that the beam density
is modulated according to the law :
\[
\rho(z)=\rho_0 +\rho_1\, \cos\tau z,\qquad \tau=\frac{2\pi}{d},
\]
where $d $ is the spatial period of modulation. Here
\[
\left.\frac{d\,^2 N_{1ph}}{d\omega
 d\Omega}\right|_{\vec{K} = \vec{\tau}}
 \]
is the spectral-angular distribution of transition radiation at
$\vec K=\vec \tau$, where $ \vec{\tau}=(0, 0,\tau)$, and $\vec
K=(\vec k_{\perp}, \frac{\omega}{v})$, where $\vec k_{\perp}$ is
the component of the photon wave vector that is orthogonal to the
$z$-axis (parallel to the boundary between the medium and vacuum).

 When the particle crosses the metal-vacuum boundary, the
spectral-angular distribution of transition radiation has the form
\cite{landau}:
\begin{equation}
\label{eq21} \frac{d\,^2N_{1ph}}{d\omega d \Omega}=\alpha\,
\frac{v^2}{\pi^2\omega\,
c^2}\frac{1}{(1-\frac{v^2}{c^2}\cos^2\vartheta_{e \,ph})^2}
\end{equation}
where $\alpha$ is the fine structure constant, $\Omega$ is the
solid angle, $\vartheta_{e\, ph}$ is the angle between the
electron velocity and the photon escape direction. The radiation
power $P=\hbar\,\omega \, \dot{N}_{ph}$, where $\dot{N}_{ph}$ is
the number of quanta emitted per unit time. The number of
electrons escaping from the SCO per unit time is defined by
formula
\[
\dot{N}_e=\frac{I}{e},
\]
where $I$ is the current of the outcoming electrons. Consequently,
\begin{equation}
\label{eq22}
\dot{N}_{ph}=\frac{I_e}{e}\left(N_{1ph}+\left.c\frac{d\,^2
N_{ph}}{d\omega
 d\Omega}\right|_{\vec K= \vec{\tau}}\,\frac{\pi}{2}\mu^2\frac{\rho_0}{K_0^2}\right),
\end{equation}
According to (\ref{eq22}), the power of coherent radiation depends
significantly on the degree of  modulation of the beam escaping
from the SCO. Since the beam continues bunching for some time
after it has escaped from the SCO, placing a foil or a wire mesh
on the beam's path at some distance from the SCO might be
beneficial for increasing the degree of beam modulation and hence
the power of coherent transition radiation.
  The power of coherent transition radiation
will increase after the beam  has passed  through such an
obstacle.

The formulas derived here provide a means to evaluate the power of
coherent radiation of a modulated beam. For example, the radiation
power for the current $I_e\approx 4-6$\,kA and the energy of
electrons from $400$ to $500$\,keV appears to be of about several
hundreds of megawatts  at frequencies in the range of several
gigahertz and the degree of beam modulation  $\mu\geq 0.5$.

An important feature of the SCO structure is that there is no need
for an applied magnetic field to guide a high-current beam. Let us
recall in this regard that similar  properties are demonstrated by
structures in which one or several electron beams pass through
periodically placed metallic threads, meshes
 (foils), or longitudinal/transverse posts forming a two- or
 three-dimensional diffraction grating (photonic crystal), being
 the  resonator of a volume free electron laser (VFEL)
 \cite{rad,HPM}.
 Volume distributed feedback that is formed in such structures not
 only provides the beam modulation and efficient generation, but also
enables self-phase locked generation  of independent beams passing
through the structure of this type \cite{VFEL}.

\section{Conclusion}
Using the method of small signal analysis, we explored the
application potential of relativistic electron beams in
split-cavity oscillators (SCO). The change in the beam energy in
SCO's as a function of the initial energy of a relativistic beam
was considered. It has been shown that the small signal analysis
method enables adequate evaluation of  SCO parameters needed for
effective modulation of a relativistic beam  in a split cavity and
for HPM generation using SCO's. It has been determined in what
energy range the  effect of self-modulation of the beam density in
SCO structures is most pronounced.
It has been demonstrated that for beam currents at which the space
charge has little effect on the electron motion in the beam, the
beam is effectively self-modulated  in SCO's  at beam energies
less than $\approx 300 \div 400$\,keV. The self-modulation drops
sharply in the range of energies from 250 to 400 keV, but as the
beam current is increased, the effective beam self-modulation
becomes appreciable in this range too, as well as  even in a
higher energy range.

\section{Acknowledgement}
I would like to thank  Sergei Anishchenko and Alexei Sytov for
valuable discussions.

\end{document}